\documentclass{article}

\usepackage{amsmath}
\usepackage{amsfonts}
\usepackage{graphicx}

\newcommand{\ud}{\mathrm{d}}

\begin{document}

\title{Theoretical framework for the emergent floe size distribution: the case for log-normality}

\author{
Fabien Montiel$^{1}$ and Nicolas Mokus$^{1}$}

\maketitle

%
%
%

\begin{abstract}
Sea ice is not continuous and homogeneous on large scales. 
Its morphology is inherently discrete and made of individual floes.
In recent years, sea ice models have incorporated this horizontal heterogeneity.
The modelling framework considers an evolution equation for the probability density function of the floe size distribution (FSD) with forcing terms that represent the effects of several physical processes.
Despite the modelling effort, a key question remains: what is the FSD emerging from the collection of all forcing processes? 
Field observations have long suggested the FSD follows a power law, but this result has not been reproduced by models or laboratory experiments. 
The theoretical framework for FSD dynamics in response to physical forcings is presented. 
Wave-induced breakup is further examined with an emphasis on how it affects the FSD.
Recent modelling results suggesting the consistent emergence of a lognormal distribution as a result of that process are further discussed.
Log-normality is also found in a dataset of floe sizes, which was originally analysed under the power law hypothesis.
We then propose a simple stochastic process of FSD dynamics, based on random fragmentation theory, that predicts log-normality. 
We therefore conjecture that the emergent FSD follows a lognormal distribution.
\end{abstract}

\section{Introduction}

Sea ice plays a key role in regulating the climate system. 
The impact of the steady decline of Arctic sea ice and the irregular, extreme patterns exhibited by Antarctic sea ice are not fully understood and sea ice models, in a fully coupled configuration, are not able to reproduce some of these patterns \cite{uotila_etal19}.
Despite much effort put into the deve-
\noindent lopment of sea ice models in recent years, the full picture of the processes governing its dynamics is not clear, in particular how to interconnect the different scales involved in sea ice dynamics, from the micro-structure of sea ice to the seasonal growth and decay on a synoptic scale.
A way forward is to shift from the large scale view of representing sea ice as a continuous and quasi-homogeneous material, to a multi-scale framework that accounts for the fact that sea ice is made of discrete floes \cite{roach_etal18b}. 
The size of floes spans multiple orders of magnitude from the sub-metre scale to at least the tenth of kilometres scale \cite{horvat_tziperman15}. 
Quantifying the floe size distribution (FSD) and understanding the processes that cause it are a key focus of modern sea ice research, especially in mixed regions such as the marginal ice zone.

Measurements of the FSD have been conducted using remote sensing techniques since the pioneering work of Rothrock \& Thorndike \cite{rothrock_thorndike84}, who analysed aerial photographs of broken sea ice in the Arctic Ocean. 
The authors described the distribution using a power-law by analysing the cumulative number distribution of mean floe calliper diameters plotted on a log-log scale.
Most subsequent observational studies have retained the power-law formalism to describe the FSDs \cite{toyota_etal06,Lu2008,steer_etal08,toyota_etal11,stern_etal18}. 
Even when clear deviations from the power-law were obtained, modified versions of the power-law were proposed, e.g.\ the truncated power-law \cite{Lu2008} or the split power-law \cite{toyota_etal11}.
Stern et al.\ \cite{Stern2018} discuss the wide range of power-law exponents reported in the literature and suggest that the widely used technique of fitting a power-law through cumulative floe size data using linear least squares in log-log space is likely to be associated with significant bias. 

The power-law formalism to describe the FSD has also trickled down to modelling work.
The study of ice-induced wave attenuation in the MIZ is often conducted by assuming the floes are distributed according to a prescribed power-law or a modified version thereof \cite{Dumont2011,Williams2013Wave-iceFoundations,montiel_etal16,squire_montiel16,williams_etal17}.
It has also been built into sea ice and FSD-evolution models \cite{zhang_etal15,Bennetts2017BriefModel}.
As the power-law parametrisation of an FSD proliferates in sea ice modelling, a key question remains: how does the observed FSD emerge from the underlying physical processes controlling it?

Toyota et al.\ \cite{toyota_etal11} attempted to address this question by conceptualising the repeated fragmentation of ice floes as a fractal process, which leads to scale invariance and therefore a power-law. 
It is unclear what would induce such a fragmentation pattern, however, as the random nature of the processes causing fracture events suggest the fragmentation process would be more random.
Recent evidence from field measurements \cite{Herman2021,dumas-lefebvre_dumont22}, laboratory experiments \cite{Herman2018,passerotti_etal22} and model simulations \cite{herman17,montiel_squire17}, further suggest that the FSD is not monotonically decreasing and instead has a clearly defined mode with two tails.
The power-law is commonly used throughout science to describe heavy-tailed empirical relationships between quantities and statistical distributions \cite{polizzi_etal21}, but upon close inspection few only seem to be statistically justified \cite{stumpf_porter12}, while a good non-monotonic alternative is the lognormal distribution \cite{Clauset2009,polizzi_etal21}.
Is the lognormal distribution a reasonable alternative for describing the FSD?

In this paper, we review the general framework for modelling the dynamics of the FSD, first introduced by Horvat \& Tzipermann \cite{horvat_tziperman15}.
The model does not make a priori assumptions on the parametric form of the FSD.
The probability density function associated with the FSD is evolved according to a transport equation with source terms parametrised to describe the different physical processes affecting the FSD (see \textsection\ref{sec2}).
These are (i) thermodynamical, including lateral growth/melt, ice formation and welding, (ii) mechanical, including rafting and ridging, and (iii) wave-induced, i.e.\ the breakup resulting from flexural gravity waves travelling through the ice cover (see \textsection\ref{sec3}).
Current parametrisations of this latter phenomenon are particularly simplified in the FSD modelling framework, and so in \textsection\ref{sec:sec4} we present an overview of the theory governing the flexurally-induced breakup of sea ice, and review the literature that has attempted to describe the FSD resulting from this phenomenon.
We show that apart from a handful of simulation-based studies \cite{herman17, montiel_squire17,Mokus2021}, most descriptions of the FSD in the context of ocean waves/sea ice interactions assume the power-law formalism.

In \textsection\ref{sec5}, we make the case that the FSD should be described by a lognormal distribution, instead of a power-law.
Our conjecture is supported by multiple lines of evidence including (i) numerical simulations of repeated wave-induced breakup in a fully coupled linear hydroelastic model \cite{Mokus2021}, (ii) an analysis of floe size field data \cite{Lu2008} showing the relative performance of several parametric distributions including the power-law and the lognormal distribution, and (iii) the theoretical description of ice floe size dynamics as a stochastic exponential growth/decay process, which is exactly described by a lognormal distribution.
Although we do not expect the lognormal FSD formalism to be universal and fit perfectly all observational datasets, we demonstrate that it should be considered as a potentially superior alternative to the power-law. 
In addition, the stochastic description of the process of repeated fragmentation that leads to the emergence of the lognormal distribution could potentially be incorporated in the FSD modelling framework discussed in \textsection\ref{sec2}, as has been done in other disciplines, e.g.\ grain size distribution in rock fragmentation \cite{fowler_scheu16}, size distribution in plant population \cite{hara84} or sea ice thickness distribution \cite{toppaladoddi_wettlaufer15}.
Some concluding remarks are given in \textsection\ref{sec6}.


\section{Theoretical framework}
\label{sec2}


Let us denote the size of a sea ice floe by $L$, which is a measure of the floe horizontal extent and has the unit of a length. 
Although several definitions of $L$ have been proposed, e.g.\ perimeter- or area-based, and assuming a disk shape \cite{toyota_etal11}, the framework described here does not depend on the choice of definition. 
Floe size $\{L_t(\mathbf{x}) \in \mathbb{R}_{+}|\mathbf{x}\in D,t\in\mathbb{R}\}$ is treated as a spatio-temporal random process, which is assumed to be weakly stationary and ergodic.  
Here $t$ denotes time and $\mathbf{x}\in D\subset\mathbb{R}^2$ is the vector of spatial coordinates, with $D$ denoting the spatial domain spanned by the sea ice cover (in flat Euclidean space).
Although the stationarity and ergodicity assumptions are reasonable based on observational measurements of floe size \cite{toyota_etal11,horvat_etal19}, they have not been rigorously tested. 
We denote the probability density function (pdf) associated with $L_t(\mathbf{x})$ by $f_L$ and refer to it as the floe size distribution (FSD).
Weak stationarity and ergodicity imply that $f_L$ is a smooth function of both $t$ and $\mathbf{x}$, and can therefore be considered a state variable of the sea ice (dynamical) system.
In the context of observations or modelling, $f_L(l)$ characterises the FSD over a region much larger than the typical floe size, e.g.\ a model grid cell, centred at $\mathbf{x}$.
It is also useful to think of $f_L(l)\ud l$ as the probability a floe of size $L$ falls in the infinitesimal interval $[l,l+\ud l]$, or equivalently as the fraction of ocean surface area covered by floes of size between $l$ and $l+\ud l$.

We further define the floe number probability density function 
\begin{equation}
f_N(l) = \frac{f_L(l)}{\mathcal{A}(l)},
\label{eq:floe_num}
\end{equation}
where $\mathcal{A}(l)$ is the surface area of a floe of size $l$, so that $\mathcal{A}(l) = k l^2$ for some constant $k$.
For instance, if all floes are assumed to be circular so that $l$
denotes the diameter of a floe then $k=\pi/4$.
Based on field observations, \cite{rothrock_thorndike84} found that a value of $k=0.66 \pm 0.05<\pi/4$ represents the area function better when floe size $l$ is defined as the mean calliper diameter.
It follows from the previous interpretation of $f_L$ that $f_N(l)\ud l$ can be thought of as the number of floes per unit ocean surface area with floe size between $l$ and $l+\ud l$.

Extending the framework for modelling the evolution of the ice thickness distribution (ITD) proposed by \cite{thorndike_etal75}, Horvat \& Tziperman \cite{horvat_tziperman15} described the transport of the joint floe size and thickness distribution (FSTD).
Restricting the present exposition to the FSD, we assume here a sea ice cover of constant thickness, $h$ say, so that the FSD $f_L$ satisfies a conservation law, which is given in differential form by
\begin{equation}
    \frac{\partial f_L}{\partial t} + \nabla . (\mathbf{F}(f_L)) = S(f_L,t),
    \label{eq:transp1}
\end{equation}
where $\mathbf{F}(f_L)$ is the convective flux of the FSD which is governed by the horizontal ice velocity field $\mathbf{v}(t,\mathbf{x})$, so that $\mathbf{F}=f_L\mathbf{v}$.
Diffusive effects in the flux vector are neglected.
In \eqref{eq:transp1}, $S(f_L,t)$ denotes the source term describing the rate at which floes are created, destroyed or redistributed in different floe size classes.
Recent modelling efforts have focused on developing parametrisations for $S(f_L,t)$ that describe sub-grid scale physical processes governing the evolution.
These are discussed in the next section.

\section{Process-informed parametrisations}
\label{sec3}

Following \cite{horvat_tziperman15} and \cite{roach_etal18b}, the source term in \eqref{eq:transp1} is written as 
\begin{equation}
    S(f_L,t) = S_T(f_L,t) + S_C(f_L,t) + S_W(f_L,t),
    \label{eq:source_terms}
\end{equation}
where each term on the right-hand side represents the contribution of a particular physical process to the rate of floe size change.
The operator $S_T$ characterises floe size changes caused by thermal effects, i.e.\ lateral growth/melt and formation of new ice.
The second term $S_C$ describes floe collisions that result in rafting, ridging or welding.
Finally, the contribution of ocean wave-induced breakup to changes in floe sizes is governed by the operator $S_W$.

\subsection{Thermal and collisional effects}

Given the seasonal cycle of growth and decay of the sea ice cover in both polar oceans, thermal processes are an obvious key driver of FSD dynamics.
Following \cite{horvat_tziperman15,roach_etal18b}, the source term $S_T(f_L,t)=S_T^{\mathrm{ex}}(f_L,t)+S_T^{\mathrm{new}}(f_L,t)$ can be decomposed into (i) lateral growth/melt of existing ice floes, denoted by $S_T^{\mathrm{ex}}$ and (ii) formation of new ice, denoted by $S_T^{\mathrm{new}}$.
We note that Roach et al.\ \cite{roach_etal18b} added a third term to $S_T(f_L,t)$ representing welding of individual floes to form larger floes as a source term.
However, since this process results from floe-floe collisions, like the mechanical interactions causing rafting and ridging, we incorporate welding into the source term $S_C$ in \eqref{eq:source_terms}.

To derive an expression for $S_T^{\mathrm{ex}}$, we invoke the conservation of floe number, assuming lateral melt/freezing is the only process affecting floe size. 
It can be written as
\begin{equation}
    \frac{\partial F_N}{\partial t} + G_l \frac{\partial F_N}{\partial l}=0,
    \label{eq:conserv_floe_num}
\end{equation}
where $F_N(l,t)=\int_0^l f_N(l',t)\ud l'$ is the cumulative floe number distribution function and $G_l$ is the lateral melt/freezing rate of a floe with size $l$ \cite{horvat_tziperman15}.
Using \eqref{eq:floe_num} in the definition of $F_N$ and differentiating \eqref{eq:conserv_floe_num} with respect to $l$, we obtain the contribution from lateral melt/freezing to floe size changes, i.e.
\begin{equation}
S_T^{\mathrm{ex}}(f_L) \equiv \left.\frac{\partial f_L}{\partial t}\right|_{\mathrm{melt/freeze}} = - \frac{\partial}{\partial l}\left(G_l f_L\right) + \frac{2}{l}G_l f_L.
\label{eq:source_therm_ex}
\end{equation}
The lateral melt/freezing rate $G_l$ is determined from the heat budget at the ocean-ice-atmosphere interface.
Following \cite{horvat_tziperman15,roach_etal18b}, an annular region around each floe is assumed to represent the region of freeze/melt in terms of the air-sea heat flux, which is partitioned into lateral and basal growth/decay. 
We note that the full FSTD is required to derive $G_l$ in \cite{horvat_tziperman15,roach_etal18b} as part of the heat flux in the lead region causes freezing/melt on the basal surface of the floes.
However, an assumption could be made to fully allocate the available heat to melting/freezing the lateral surface of the floes alone, i.e.\ with no change in thickness, therefore allowing us to be consistent with the present FSD dynamics framework.

To account for new ice formation from the air-sea heat flux, we can assume that newly-created ice must have a prescribed minimum floe size $L_{min}$ and the same thickness $h$ as all existing floes.
Typically, $L_{min}$ represents the size of pancake ice as the minimum floe size resolved in the current framework \cite{horvat_tziperman15,roach_etal18b}.
With this assumption, the source term for sea ice formation is given by
\begin{equation}
    S_T^{\mathrm{new}} \equiv \delta(l-L_{min})\Dot{A}_p,
    \label{eq:source_therm_new}
\end{equation}
where $\Dot{A}_p$ is the time rate of area growth of pancake ice, which can be parametrised in terms of the air-sea heat flux and the open water area available for new ice to form \cite{horvat_tziperman15,roach_etal18b}, and $\delta(.)$ denotes the Dirac delta function.
We note that the choice of minimum floe size $L_{min}$ plays an important role in the resulting simulated FSD, as shown by Roach et al.\ \cite{roach_etal18b}.
In a subsequent paper \cite{roach_etal19}, these authors were able to refine the parametrisation of ice formation, by deriving a wave-dependent pancake ice floe size in the presence of waves of sufficiently energetic waves. 
In a calm sea, however, frazil crystals formed at the free surface consolidate into nilas ice (thin sheet of sea ice).
This process was also considered in \cite{roach_etal19}.
We note that our assumption of constant thickness would clearly not be compatible with the phenomenon of nilas ice growth, and so we do not consider this ice type further in this paper. 

Floe-floe collisions are another important driver of FSD changes.
For instance, under freezing conditions, floes that randomly collide can weld to each other to form larger floes.
Under the assumption of ice area conservation, welding was first incorporated into the present modelling framework by Roach et al.\ \cite{roach_etal18b}, following field observations of this process \cite{roach_etal18a}.
Collisions between floes can also lead to rafting and ridging. 
As opposed to welding, these processes cause a decrease in ice area and therefore an increase in open water area, assuming conservation of ice volume. 
Horvat and Tziperman \cite{horvat_tziperman15} parametrised these mechanical interactions into \eqref{eq:transp1} via the source term $S_C(f_L,t)$, but did not consider welding.

Changes to the FSD resulting from floe-floe collisions are modelled using the coagulation theory of Smoluchowski \cite{smoluchowski_16}, which describes, under a probabilistic framework, the time evolution of the number density function of particle size (in our case particles are floes).
Specifically, the floe number distribution satisfies
\begin{equation}
    \frac{\partial f_N}{\partial t} = \frac12\int_0^l\int_0^l K(l',l'';l) f_N(l') f_N(l'')\,\mathrm{d}l'\mathrm{d}l'' - \int_0^\infty \int_0^\infty K(l,l'';l') f_N(l) f_N(l'')\,\mathrm{d}l'\mathrm{d}l'',
    \label{eq:coag}
\end{equation}
where the integral kernel $K(l',l'';l)$ describes the rate at which floes of size $l'$ and $l''$ coagulate to form floes of size $l$.
Therefore, it must satisfy $K(l',l'';l)=0$ if $l'+l''>l$.
The exact parametrisation of the kernel depends on the physical process considered.
For instance, welding of floes is assumed to preserve ice area, so the corresponding coagulation equation \eqref{eq:coag} and associated kernel are more naturally expressed after making the change of variable $\mathcal{A}=kl^2$ (see \cite{roach_etal18b}).
These authors then derive a simplified evolution equation for the corresponding number distribution $\tilde{f}_N(\mathcal{A})$, i.e.\ $\partial_t \tilde{f}_N = -\kappa c^2/2$, where $\kappa$ is the prescribed rate of welding per unit area, per unit time, and $c$ is ice concentration.
The reader is referred to \cite{roach_etal18b} for the full derivation.

A different kernel parametrisation is used to model rafting and ridging under a convergent ice flow regime, i.e.\ $\nabla.\mathbf{v}<0$, which yields to a decrease in ice area and therefore an increase in open water area.
Assuming conservation of ice volume, Horvat and Tziperman \cite{horvat_tziperman15} proposed kernel parametrisations for these two processes, noting that the ice area loss is balanced by an increase in ice thickness.
Again, the reader is referred to the original paper for a full derivation.
Given the collective impact of all these collision-dependent processes, the corresponding source term in \eqref{eq:transp1} is given by
\begin{equation}
    S_C(f_L,t) = \left.\frac{\partial f_L}{\partial t}\right|_{\mathrm{coll}} = k l^2 \frac{\partial f_N}{\partial t},
\end{equation}
where $\partial_t f_N$ is given by \eqref{eq:coag}.


\subsection{Wave-induced breakup}

In addition to affecting the formation of sea ice into pancake ice floes, ocean waves have been observed to bend the ice they travel through so that if the flexural stress exceeds locally the material flexural strength, a fracture occurs resulting in the creation of two smaller floes.
This process has been parametrised in the FSD dynamics modelling framework presented here by Horvat and Tziperman \cite{horvat_tziperman15} and Zhang et al.\ \cite{zhang_etal15}, under the reasonable assumption that sea ice area is preserved, i.e.\ 
\begin{equation}
    \int_0^\infty S_W(f_L(l),t)\,\mathrm{d}l=0.
    \label{eq:area_conserv_wave}
\end{equation}

Let $\Omega(l,t)$ define the fractional density of ocean surface covered by floes of size $l$ that breaks up under wave action per unit time.
As a result of wave-induced breakup, a fraction of all floes of a given size $l$ is redistributed into smaller floe sizes.
This process can be modelled as
\begin{equation}
    S_W(f_L(l),t) = - \Omega(l,t) + \int_l^{\infty} K_w(l,s) \Omega(s,t)\,\mathrm{d}s,
    \label{eq:source_wave}
\end{equation}
where the integral kernel $K_w(l,s)$ describes the fraction of ocean surface covered by floes of size $l$ created from the breakup of floes of size $s$, so that $s>l$.
The first term in \eqref{eq:source_wave} is the fractional area of floes with size $l$ that is lost due to wave fracture, while the second term corresponds to the fractional area with size $l$ that is gained from the breakup of larger floes.
The area conservation equation \eqref{eq:area_conserv_wave} requires that $\int_0^\infty K_w(l,s) \,\mathrm{d}l=1.$

Following Horvat and Tziperman \cite{horvat_tziperman15}, parametrisations for $\Omega(l,t)$ and $K_w(l,s)$ are derived by prescribing a post-breakup floe size number distribution function $D_w(l)$.
We note that in their paper, Horvat and Tziperman refer to $D_w(l)$ as a histogram, therefore assuming floe size $l$ can only exist in a predefined number of discrete bins, even though their formulation is also valid for a continuous number distribution function. 
The distribution $D_w(l)$ is empirically determined via observation or wave-ice interaction model outputs. 
The scheme used by \cite{horvat_tziperman15}, and also by \cite{roach_etal18b}, will be described in \textsection\,\ref{sec:sec4}, along with other possible approaches that have not yet been incorporated in the FSD dynamics framework discussed here.

Once equipped with a post-breakup floe size number distribution $D_w(l)$, the integral kernel in \eqref{eq:source_wave} can be computed as the proportion of broken floes with pre-breakup size $s$ that have post-breakup size $l$, i.e. 
\begin{equation}
    K_w(l,s) = \frac{lD_w(l)}{\int_0^s l'D_w(l')\,\mathrm{d}l'}H(s-l),
    \label{eq:kernel_wave}
\end{equation}
where $H(.)$ is the Heaviside step function, ensuring that $K_w(l,s)=0$ if $l>s$.
Analogously, the fractional area of floes with pre-breakup size $l$ that experience wave-induced breakup per unit time is given by
\begin{equation}
    \Omega(l) = \frac1{t_w}\mathrm{P}_{br}(l)f_L(l),
    \label{eq:floe_break_wave}
\end{equation}
where $t_w$ is a time scale describing the breakup event that generated $D_w(l)$, while the probability a floe of size $l$ experiences wave-induced breakup is given by $\mathrm{P}_{br}(l) = \left(\int_{0}^l l'D_w(l')\,\mathrm{d}l'\right)/\left(\int_{0}^{\infty} l'D_w(l')\,\mathrm{d}l'\right)$.
We note that the time scale $t_w$ was estimated by Horvat and Tziperman \cite{horvat_tziperman15} to be the ratio of the group velocity associated with the mean zero-crossing wavelength and the total extent of broken floes, i.e. $t_w=\left(\int_{0}^{\infty} l'D_w(l')\,\mathrm{d}l'\right)/c_g$.

The model describing how wave-induced sea ice breakup affects the dynamics of the FSD, summarised by \eqref{eq:source_wave}--\eqref{eq:floe_break_wave}, depends entirely on the parametrisation chosen for the post-breakup floe size number distribution function $D_w(l)$.
Unfortunately, there is no consensus in the wave-ice modellers community on how to model this phenomenon appropriately. 
In addition, field and laboratory observations of wave-induced sea ice breakup are scarce and not sufficient to derive an empirical model for $D_w(l)$.
In the next section, we summarise the literature that has so far attempted to answer the key question: what is the FSD generated by wave-induced sea ice breakup?

\section{Theory of wave-induced sea ice breakup}
\label{sec:sec4}

\subsection{Fracture mechanics}
\label{sec:sec4.1}

Although sea ice is composed of discrete floes, on a scale much larger than individual floes, it is useful to model its dynamics as that of a deformable continuum material.
Sea ice rheology, describing the relationship between the stress and strain rate tensors of the continuum, has been the subject of extensive investigation, particularly since the seminal Arctic Ice Dynamics Joint Experiment (AIDJEX) \cite{feltham08}.
Most modern sea ice modules in earth system models are based on the viscous-plastic rheology of Hibler \cite{hibler79} or modifications thereof that include elastic deformations to describe the horizontal deformation of the sea ice cover.
Recently, elasto-brittle rheology was considered in a number of large scale models to represent the formation and evolution of cracks and leads \cite{girard_etal11,bouillon_rampal15,dansereau_etal16}.

The deformations caused by ocean waves as they travel through an ice-covered ocean are different, as sea ice mainly experiences flexural motion which is the result of a non-uniform horizontal stress distribution across the ice thickness.
The relevant spatial scale over which these deformations take place ($O(10\text{--}10^2)$\,m) is much smaller than the typical grid size of a large scale sea ice model, so that the ice rheology of the continuum ice modelling framework mentioned above is not appropriate to describe the wave-induced flexural response of sea ice.
At such a small scale, the mechanical properties of individual ice floes, which can differ significantly from the effective mechanical properties of a continuous sea ice cover, can be measured directly \cite{timco_weeks10}.
In characterising the flexural motion of sea ice floes under wave action, it is well accepted that floes can be modelled as elastic plates, typically under the paradigm of linear elasticity and the plane stress approximation. 

The most common framework for modelling the flexural motion of sea ice floes is the Kirchhoff-Love (KL) thin-elastic plate theory.
The model relies on many simplifying assumptions \cite{reddy06}, not the least of which is that sea ice is a homogeneous solid material.
In reality, sea ice is inhomogeneous and even porous, with a complex anisotropic microstructure that depends on the way it is formed \cite{timco_weeks10}.
Its mechanical properties depend on many factors, including temperature, brine volume fraction and also its dynamical history, e.g.\ fatigue \cite{timco_weeks10}. 
We refer the reader to \cite{squire_montiel15} for an overview of how these factors can be incorporated into the KL theoretical framework.
Under this theory, the three non-zero components of the stress and strain tensors in Cartesian coordinates are related via Hooke's law 
\begin{equation}
    \left(\begin{array}{c}
         \sigma_{xx}  \\ \sigma_{xy} \\ \sigma_{yy}
    \end{array}\right) = \frac{E}{1-\nu^2}
    \left(\begin{array}{ccc}
         1 & \nu & 0  \\ \nu & 1 & 0 \\ 0 & 0 & 1-\nu
    \end{array}\right)
    \left(\begin{array}{c}
         \epsilon_{xx}  \\ \epsilon_{xy} \\ \epsilon_{yy}
    \end{array}\right),
    \label{eq:strain2stress}
\end{equation}
where $\sigma_{ii}$ and $\epsilon_{ii}$ ($i=x,y$) are the normal stress and strain, respectively, in the two horizontal dimensions, while $\sigma_{xy}$ and $\epsilon_{xy}$ denote the shear stress and strain, respectively.
We have also introduced the effective Young's modulus $E$ and Poison's ratio $\nu\approx0.3$, which we assume to be constant.
This constitutive relationship assumes a purely elastic flexural behaviour.
Different constitutive laws, for instance incorporating viscous effects, can also be used (see, e.g., \cite{montiel_etal13b}), but for the purpose of the present discussion, we assume the plate's flexural deformations are purely elastic.
The components of the strain tensor can be further expressed in terms of the vertical deflection of the plate's middle surface, $\eta(\mathbf{x},t)$ say, with respect to its position at rest, as
\begin{equation}
    \epsilon_{xx} = -z \frac{\partial^2\eta}{\partial x^2}, \quad \epsilon_{xy} = -z \frac{\partial^2\eta}{\partial x\partial y}, \quad \epsilon_{yy} = -z \frac{\partial^2\eta}{\partial y^2},
    \label{eq:def2strain}
\end{equation}
where $z$ is the vertical coordinate pointing upward, which vanishes at the plate's middle surface at rest.
We see that if the deflection field of a floe $\eta(\mathbf{x},t)$ is known, then \eqref{eq:def2strain} and \eqref{eq:strain2stress} fully determine the strain and stress fields throughout the elastic plate.
Given the geometry of a floe and a surrounding linear wave field, wave scattering theory and linear hydroelasticity can be used to derive the deflection field \cite{squire07,squire11}.
The key question remains: under which conditions does a floe break?

Under the assumption of purely elastic deformations and homogeneous material properties, a fracture occurs when the components of the stress tensor locally combine to exceed the material strength. 
For a thin-elastic plate, this condition takes the form 
\begin{equation}
    \widetilde{\mathcal{F}}(\sigma_{xx},\sigma_{xy},\sigma_{yy})=0,
    \label{eq:fail_surf3}
\end{equation}
which can be interpreted as a surface in the three-dimensional stress space \cite{gross_seelig11}. 
Material strength can differ under different types of loading, e.g.\ compressive, tensile, shear or flexural, and the strength of sea ice has been measured under these different forcing \cite{timco_weeks10}.
However, both shear and flexural stresses are biaxial stress conditions composed of tensile and compressive stresses only.
Therefore, the three-dimensional failure surface \eqref{eq:fail_surf3} can be reduced to a failure curve in a two-dimensional stress space with coordinates $\sigma_1$ and $\sigma_2$, which are the eigenvalues of the stress tensor expressed as a symmetric 2 by 2 matrix with diagonal components $\sigma_{xx}$ and $\sigma_{yy}$ and off-diagonal component $\sigma_{xy}$.
The resulting failure curve 
\begin{equation}
    \mathcal{F}(\sigma_1,\sigma_2) = 0,
\end{equation}
is referred to as the yield curve.

\begin{figure}
    \centering
    \includegraphics[width=0.7\textwidth]{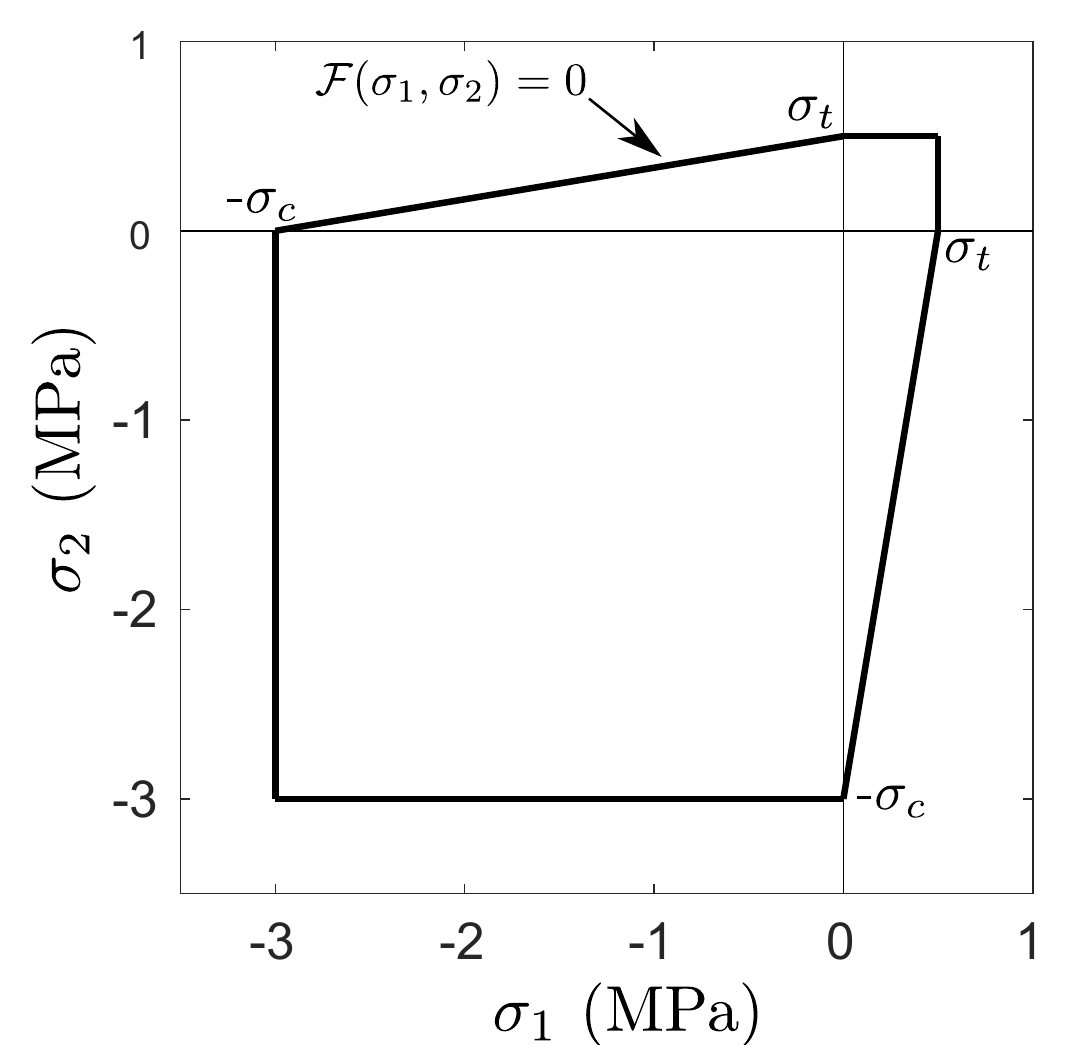}
    \caption{MC yield curve for a sea ice thin elastic plate with tensile strength $\sigma_t=0.5$\,MPa and compressive strength $\sigma_c=3$\,MPa (adapted from Montiel \& Squire \cite{montiel_squire17}).}
    \label{fig1}
\end{figure}

Determining experimentally the yield curve of a material is difficult, especially for a material like sea ice, as combinations of compressive and tensile loads must be applied and failure be measured in response.
Timco \& Weeks \cite{timco_weeks10} report some attempts to measure the yield curve of sea ice under various conditions, but do not provide empirical descriptions for practical use in modelling wave-induced sea ice breakup.
On the other hand, the sea ice strength under uni-axial loads is fairly well understood, with tensile strength $\sigma_t=O(0.1)$\,MPa being typically one order of magnitude smaller than compressive strength $\sigma_c=O(1)$\,MPa \cite{timco_weeks10}.
From that perspective, it is sensible to use idealised parametrisations of the yield curve in terms of $\sigma_t$ and $\sigma_c$ only.
One such choice is the Mohr-Coulomb failure hypothesis \cite{gross_seelig11}, which can be expressed in the case of a thin-elastic plate as
\begin{equation}
    \mathcal{F}(\sigma_1,\sigma_2) \equiv \sigma^{(\mathrm{max})}(\sigma_1,\sigma_2) - \sigma^{\mathrm{MC}} = 0, 
    \label{eq:MC_yield}
\end{equation}
where
\begin{equation}
    \sigma^{(\mathrm{max})}(\sigma_1,\sigma_2) = \mathrm{max}\left\{
    \left|\sigma_1-\sigma_2\right| + K(\sigma_1+\sigma_2),
    \left|\sigma_1\right| + K\sigma_1,\left|\sigma_2\right| + K\sigma_2\right\},
    \label{eq:sig_max}
\end{equation}
with
\begin{equation}
    K = \frac{\sigma_c-\sigma_t}{\sigma_c+\sigma_t} \quad \text{and} \quad \sigma^{\mathrm{MC}} = \frac{2\sigma_c\sigma_t}{\sigma_c+\sigma_t}.
\end{equation}
An example of the MC yield curve is shown in figure \ref{fig1}.
We note the curve is symmetric with respect to the axis $\sigma_1=\sigma_2$.
Based on \eqref{eq:MC_yield}, a fracture occurs locally in the plate when $\sigma^{(\mathrm{max})}(\sigma_1,\sigma_2) \ge \sigma^{\mathrm{MC}}$. 
Extensions to the MC yield curve are possible, e.g.\ the Drucker-Prager hypothesis \cite{gross_seelig11}, but to our knowledge, they have not been used to model sea ice fracture.

\subsection{Floe size distribution}
\label{sec4.2}

\subsubsection{Non-parametric descriptions}

The parametrisation of wave-induced sea ice breakup introduced in \textsection\ref{sec3} describes its influence on the FSD on large scales.
It requires a post-breakup floe size number distribution $D_w(l)$, which can be determined in a number of ways.
Given the fracture mechanics framework discussed in \textsection\ref{sec:sec4}.\ref{sec:sec4.1}, which describes how sea ice breakup can be parametrised on small scales, an intermediate model is needed to bridge the two scales.
This is typically provided by a hydroelastic model of ocean waves/sea ice interactions, which predicts the deflection field $\eta(\mathbf{x},t)$ experienced by an array of floes under linear wave forcing \cite{squire07,squire11}.
Then \eqref{eq:def2strain} and \eqref{eq:strain2stress} can be used to recover the strain and stress fields.

The MC fracture criterion discussed in \textsection\ref{sec:sec4}.\ref{sec:sec4.1} was used by Montiel \& Squire \cite{montiel_squire17} to model the repeated breakup of circular, thin elastic ice floes.
The model is based on multiple scattering theory for large random arrays of circular floes \cite{montiel_etal15,montiel_etal16,squire_montiel16} to compute the stress field experienced by each floe under monochromatic plane wave forcing.
The breakup criterion for each floe is then defined as $\sigma_{\mathrm{br}} \ge \sigma^{\mathrm{MC}}$, where $\sigma_{\mathrm{br}}$ is the maximum value of $\sigma^{(\mathrm{max})}(\mathbf{x},t)$, as defined in \eqref{eq:sig_max}, over the surface of a floe and over one period cycle of the forcing wave.
Simulations were then conducted to repeatedly fracture an initial ice cover under time-harmonic wave forcing until a steady state is reached.
Although a large parameter space was explored, the authors consistently found that the resulting post-breakup FSD, i.e.\ $D_w(l)$, was uni-modal with limited support and nearly symmetric, suggesting near normality.
We note that no attempt was made to fit parametrised distribution in that study, because the sample size of floe sizes post-breakup was too small to ensure statistical significance.

As far as the authors are aware, \cite{montiel_squire17} is the only study that has modelled wave-induced fracture of ice floes in three dimensions, where floe dynamics is described by thin-elastic plate theory.
Even in two dimensions, there is a relative paucity of published studies that have attempted to model wave-induced breakup with the goal of estimating the resulting FSD.
In fact, only Herman \cite{herman17} attempted this exercise, without assuming the FSD follows a specific functional form. 
This author coupled a nonlinear wave model to a floating sea ice structure composed of rigid, massive grains connected by small elastic beams called bonds.
The wave model, i.e.\ NHWAVE \cite{ma_etal12}, is based on the incompressible, non-hydrostatic Navier-Stokes equations.
A stress-based fracture criterion, parametrised by the tensile strength of the bonds, was implemented. 
The transient response, including breakup, of the sea ice structure to both regular and irregular wave forcings was simulated. 
The author found that the distance from the ice edge at which the maximum stress and therefore breakup occurs has very little dependence on the wave frequency, therefore resulting in a very narrow FSD with floes that almost all have the same size.

The main reason why so little work has been conducted to quantify the FSD caused by wave-induced breakup is because of the widespread assumption within the wave modelling community that the FSD follows a power-law (Pareto distribution) or a variation thereof.
Therefore, most wave modelling studies in the MIZ, including those that incorporate wave-induced breakup, assume the FSD follows the parametric form $f_N(l) = C l^{-\gamma}$, for some constant $C$ and exponent $\gamma>1$.

\subsubsection{The power-law formalism}
\label{sec4.2.2}
Following observations in the Antarctic MIZ, Toyota et al.~\cite{toyota_etal11} adopted a renormalisation group method to explain what they interpreted as two distinct size regimes to which they fitted Pareto distributions with different exponents; what is often called a split power-law.
This method introduces a fragility parameter from which the distribution exponent can be derived.
The fragility can be made to depend on the ice, e.g.\ thickness, or the wave conditions, e.g.\ significant wave height.
Subsequently, Dumont et al.~\cite{Dumont2011} proposed an ice--wave interaction model, with the aim of improving wave forecasting in polar regions.
This model introduces a two-way coupling where the wave field, advected into the ice, is attenuated and at the same time is likely to transform the FSD by fracturing floes.
They assumed translational invariance in one horizontal direction, reducing the strain tensor to a scalar field.
Breakup is triggered when the propagating wave, assumed to generate sinusoidal perturbations in the ice cover, causes a deflection sufficient for the strain \eqref{eq:def2strain} to exceed an experimentally determined strain limit.
As the linear water wave theory underpinning this definition of flexural strain does not allow for cavitation under the ice plate, they doubled this failure criterion with another parametrisation, based on the stress field and the ice flexural strength, which takes it into account.
This yields two different amplitude thresholds leading to wave-induced fracture, both depending on the propagating wavelength and the ice thickness.
The attenuation of the wave amplitude is based on the mean floe size, which is derived by assuming a parametric form of the FSD.
The authors used the formalism introduced by~\cite{toyota_etal11} to justify the use of a Pareto distribution, but introduced an upper truncation term, equivalent to the largest floe size admissible.
As a sinusoidal wave causes strain extrema separated by half its wavelength, when breakup occurs, the upper limit is set to half the propagating wavelength, effectively impacting the derived mean floe size.
This model is extended by~\cite{Williams2013Wave-iceFoundations}, with the addition of viscous damping, and a simplification of the breakup scheme which no longer takes the flexural strength into account and relies on strain failure exclusively.
They allow for the propagation of a wave spectrum, enabling constructive interference to affect the strain undergone by the ice.
This model constitutes the backbone of the waves-in-ice model coupled to the sea ice model neXtSIM~\cite{williams_etal17}.
This study was, again, concerned with the impact of the ice on the wave field rather than on the shaping of the FSD.

These models~\cite{Dumont2011, Williams2013Wave-iceFoundations} were generalised by Boutin et al.~\cite{Boutin2018FloeEvaluation} as a component of the spectral model WAVEWATCH III (WW3).
This study was focused on improving the representation of the wave attenuation by an ice cover.
While highlighting the importance of the feedbacks between waves and sea ice, they focused on improving the representation of the wave attenuation by an ice cover and assume the breakup-driven FSD follows a Pareto distribution.
Bennetts et al. 2017~\cite{Bennetts2017BriefModel} implemented a wave-induced breakup parametrisation similar to that of \cite{Williams2013Wave-iceFoundations} in the sea ice model CICE, in order to study the impact of wave-induced breakup on lateral melt.
The sub-grid FSD was assumed to follow the split power-law suggested by~\cite{toyota_etal11}.
Sharing similar goals, Bateson et al.~\cite{Bateson2020ImpactIce} implemented in CICE a waves-in-ice module to study the effect of introducing the FSD on sea ice extent and volume.
Their breakup parametrisation was inspired by that of~\cite{Williams2013Wave-iceFoundations} and assumed an underlying Pareto distribution.
Building further on their WW3 module, Boutin and colleagues coupled this it to the sea ice models LIM3~\cite{Boutin2018FloeEvaluation} and neXtSIM~\cite{Boutin2021}.
Even though the FSDs are evolved by the ice models, the wave-induced breakup is handled by WW3, which, for coherence, leads to convergence towards Pareto distributions.

\subsubsection{Limitations and recent evidence}
\label{sec4.2.3}

Most authors cited in \textsection\ref{sec:sec4}.\ref{sec4.2}.\ref{sec4.2.2} acknowledged that settling on the Pareto distribution has some influence on their results, but studying the FSD itself was not their goal, unlike~\cite{montiel_squire17, herman17} who specifically focused on the FSD emergence under wave action.
As discussed earlier, neither of these studies replicated the Pareto distributions.
Moreover, although the Pareto distribution formalism emerged from observational data, the exponents fitted to observed FSDs vary largely depending on the scales considered, the time of the year, the proximity of the ice edge, and the location~\cite{Stern2018}, suggesting that the "one size fits all" power-law distribution adopted by most of the aforementioned studies may not be able to capture this diversity.
Additionally, it is not clear how wave action itself, and not other processes, would cause such a distribution.
In a laboratory experiment of sea ice breakup caused by waves, Herman et al.~\cite{Herman2018} found distributions qualitatively similar to the ones observed by~\cite{toyota_etal11}, which they parametrised as combinations of a normal distribution and a gamma distribution; an approach later used successfully on field observations~\cite{Herman2021}.
The normal distribution is hypothesized to capture the FSD caused by the breakup from waves, while the gamma distribution is the result of all other FSD altering processes.
Recent observational results (though not yet published) attempting to control the wave conditions leading to breakup in both field and laboratory settings seem to support the FSD resulting from this process is uni-modal and possibly skewed~\cite{dumas-lefebvre_dumont22,passerotti_etal22}.

In a recent (yet to be published) study by the current authors (Mokus and Montiel~\cite{Mokus2021}), we traded the higher order physics of~\cite{herman17} and the 3D-geometry of~\cite{montiel_squire17} for computational efficiency, allowing for the treatment of large numbers of floes.
A semi-infinite floating elastic plate was repeatedly broken under monochromatic wave forcing.
The resulting FSDs were shown to be consistently well fitted by a lognormal distribution, consistent with the uni-modality and skewness seen in observational studies.
This result was extended to the breakup caused by a wave spectrum for a range of wave and ice conditions.

\section{Emergent FSD}
\label{sec5}

Recent evidence discussed in the previous section seem to challenge the long-standing conjecture that the FSD follows a power-law/Pareto distribution.
This is not unique to the FSD.
Power laws are often fitted to datasets, because of the apparent linearity in the data when visualised on logarithmic scales.
Causet et al.~\cite{Clauset2009} tested the validity of the power law used in 14 empirical studies and found that it often failed statistical scrutiny, while also showing the lognormal distribution performs better following a relative goodness of fit test.  
The choice of a lognormal distribution for these tests was not arbitrary.
Like power laws, lognormal distributions have been used to describe a wide range of empirically-observed phenomena.
In particular, the lognormal distribution has been shown to fit well particle size distribution obtained from the fragmentation of rocks, e.g.\ as a result of milling or volcanic explosion~\cite{fowler_scheu16}.

The emergence of the lognormal distribution can be understood from a repeated fragmentation process.
This was first described by Kolmogorov~\cite{kolmogorov41}, who argued that successive fragmentation of a particle at a random position implies the relationship $d_{n+1}=f_nd_n$ between the particle size before breakup $d_n$ and after breakup $d_{n+1}$, and where $f_n$ is drawn from a random distribution on $(0,1)$.
This is referred to as a multiplicative process.
Taking the natural log yields
\[\ln d_{n+1} = \ln f_n + \ln d_n = \ln d_0 + \sum_{i=1}^n \ln f_i\]
As a sum of random samples, $\ln d_n$ tends to follow a normal distribution as a consequence of the central limit theorem.
Therefore, $d_n$ is lognormaly distributed.
Given the obvious analogy between the fragmentation process described by Kolmogorov and the breakup of sea ice under wave forcing and possibly other processes, we now attempt to test the goodness of fit of the lognormal distribution to FSD data.

\begin{figure}[ht]
    \centering
    \includegraphics[width=1\textwidth]{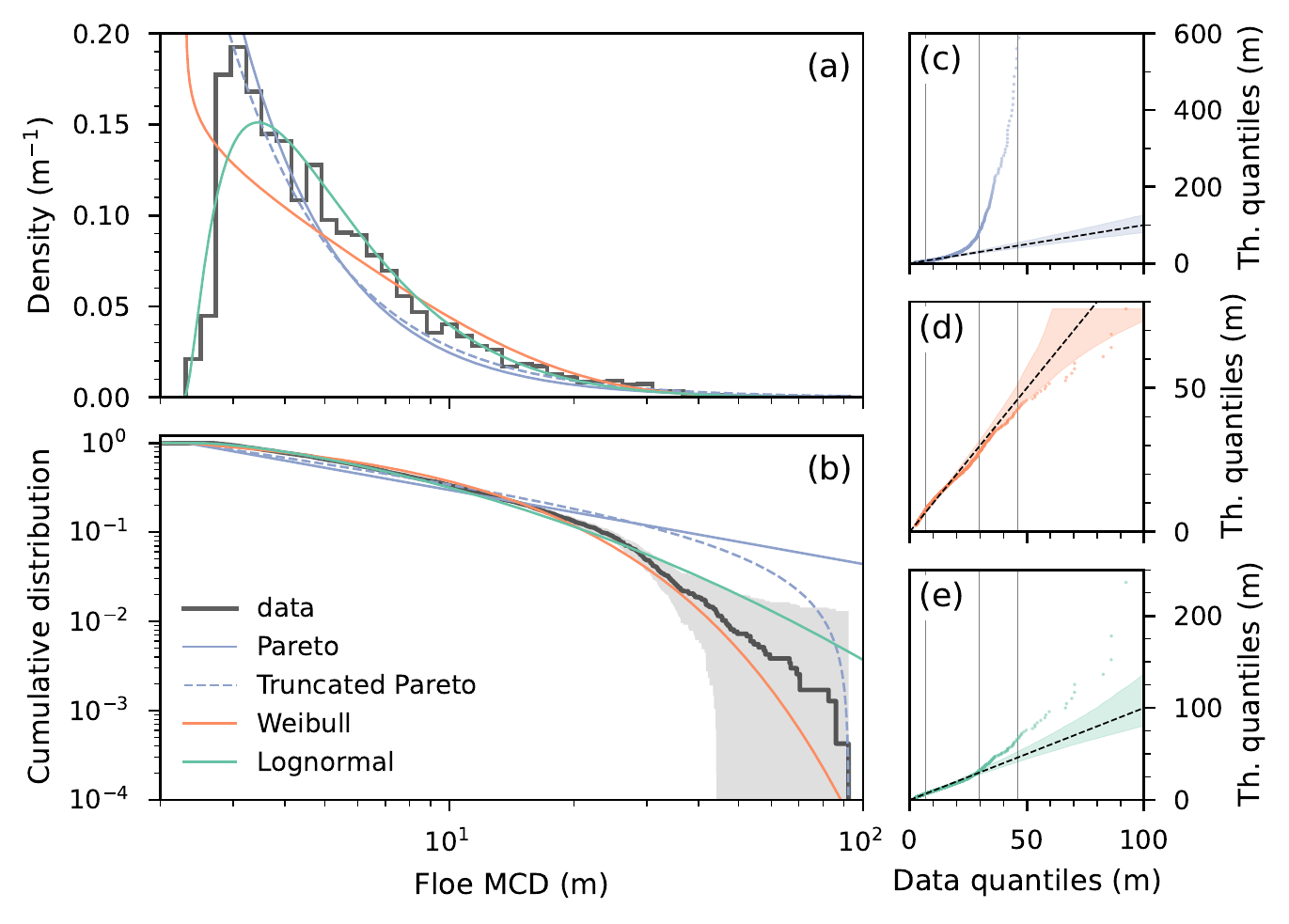}
    \caption{FSD from~\cite{Lu2008}, including 2344 data points. (a) Floe number histogram and PDFs of four different models. (b) Empirical cumulative floe number distribution and CDFs of four different models. (c) Quantile--quantile plot comparing the observations (x-axis) to the quantiles of the corresponding Pareto distribution (y-axis). The 3 vertical lines indicate, from left to right, the median, the 95th percentile and the 99th percentile of the data. (d--e) Idem as (c) for, respectively, the Weibull and the lognormal distributions. Shaded areas on (b--e) indicate 95\% confidence intervals on the empirical CDF and quantiles. For clarity, the vertical axis on (c) has been truncated to the 99th percentile of the reference Pareto distribution.}\label{fig2}
\end{figure}

The data analysed here were obtained in the Southern Ocean during the summer 2004--2005~\cite{Lu2008}.
We conduct our analysis on a strip of approximately 2000 by 200\,m, designated as Area of Interest~1, including 2344 identified ice floes.
The mean calliper diameter (MCD) is used as a metric of floe size.
The horizontal resolution is around 30~cm, ensuring the identification of small floes, but excluding large floes as only those not intersecting with image boundaries are considered.
The authors of the initial study acknowledged the limitations of the Pareto distribution model fit, and suggested two alternatives: the upper-truncated Pareto distribution and the Weibull distribution.
Here, we compare these three distributions to the lognormal.
We estimate distribution parameters by maximum likelihood and display the results on Fig.~\ref{fig2}.
Unfortunately, information on the wave conditions which might have led to the ice cover disruption are not available.
Therefore, we do not attempt to make a comparison to modelling results.

Figure~\ref{fig2}~(a) shows a comparison between a histogram of floe sizes and the fitted density functions.
The histogram is left-skewed, and the modal floe size is clearly distinct from the minimum floe size, a feature only captured by the lognormal distribution.
The density of the fitted Weibull (often referred to as stretched exponential) diverges for small floe sizes, while the densities of both the Pareto variants reach large, finite values.
From Fig.~\ref{fig2}~(b) it is clear that the power-law behaviour of the cumulative distribution does not extend to the whole range of identified floe sizes.
The lognormal and Pareto distributions tend to overpredict larger floes, while the Weibull distribution does the best job at representing the right tail behaviour.
It is even clearer from the quantile-quantile plots, shown in Fig.\ref{fig2}~(c--e), we use to visually assess the fits.
Points above the diagonal indicate over-representation by the fitted model.
The lognormal distribution represents the data fairly well (points aligned with the main diagonal) until their 95th percentile, but fails at capturing more extreme values.
We note that large floe sizes are, because of experimental constraints, absent from the dataset even though they might have existed when the data were acquired.
We present a quantitative comparison of these distributions in Table~\ref{tab:tab1}.

\begin{table}[ht]
    \centering
    \caption{Model comparisons using the likelihood ratio test. For each of the models estimated and presented on Fig.~\ref{fig2}, we quote the associated maximised log-likelihood $\ell$, ordered by decreasing $\ell$. We compare the models to the best-performing one, in this case the lognormal distribution, where a positive $\Delta\ell$ indicates failure to reject the lognormal distribution in favour of the alternative model at the confidence level $1-p$. The $p$-values being small, we quote their logarithms.}\label{tab:tab1}
    \begin{tabular}{lccc}
        \noalign{\hrule height 0.20ex}
        Model & $\ell$ & $\Delta\ell$ & $\log_{10}p$\\
        \noalign{\hrule height 0.10ex}
        lognormal &        -7161.1 &     & \\ 
        Weibull &          -7241.4 & 5.3 & -6.84 \\
        truncated Pareto & -7411.9 & 14  & -45.8 \\
        Pareto &           -7573.2 & 22  & -106 \\
        \noalign{\hrule height 0.20ex}
    \end{tabular}
\end{table}

We use the likelihood ratio test to classify the presented models.
A description of this framework can be found in~\cite{Clauset2009}.
This test assesses which one of two distributions is better suited to representing data, and whether the difference is large enough to be significant.
We only compare the lognormal distribution, which has the highest maximised likelihood, to the three others.
These comparisons point to the lognormal distribution being largely superior to the competing distributions.
These experimental evidence suggest that this model should be considered as a candidate for describing future FSD observations, but they also bring up some questions about its emergence from the underlying physical processes that describe the evolution of the FSD (see \textsection\ref{sec3}).

The simple argument for the emergence of the lognormal distribution from repeated fragmentation proposed by Kolmogorov~\cite{kolmogorov41} can be extended to a growing (as opposed to decaying) multiplicative process, i.e.\ when $f_n$ can take values greater than 1.
This is required in the context of sea ice floe size evolution, which can experience growth and decay as discussed in~\textsection\ref{sec3}.
Although that model is satisfying in explaining the emergence of a lognormal distribution, it provides little quantitative predictability and is not directly compatible with the FSD evolution modelling framework presented in \textsection\ref{sec2} and \ref{sec3}.
We propose that the latter could be achieved by deriving a time-continuous stochastic process model.
The following description closely follows and was inspired by the recent work of Anderson~\cite{andersson21} for explaining the emergence of the lognormal distribution for particle concentration in the atmosphere.

Referring back to \textsection\ref{sec2}, let us think of floe size as a time-evolving random process $L_t$. 
The simplest possible growth/decay model we can consider under this stochastic paradigm is
\begin{equation}
    \frac{\mathrm{d} L_t}{\mathrm{d} t} = -k(t) L_t,
    \label{eq:Stoc_exp_model}
\end{equation}
where the growth/decay rate $k(t)\,\in\mathbb{R}$ is evolving with time $t$ in a random fashion.
Following \cite{andersson21}, $k(t)$ can be expressed as the sum of a constant deterministic part $\mu_k$ and a stochastic part with magnitude $\sigma_k\,\in\mathbb{R}^+$, so that $k(t) = \mu_k + \sigma_k\eta(t)$, where $\eta(t)$ is a random noise parameter assumed to be independent and normally distributed (with mean 0 and standard deviation 1).
Equation \eqref{eq:Stoc_exp_model} is a stochastic differential equation (SDE) exhibiting multiplicative noise, i.e.\ the noise parameter multiplies the variable $L_t$. 
The standard formulation for SDEs relies on additive noise, however, so we introduce the change of variable $Q_t = \ln L_t$, which satisfies the Langevin equation
\begin{equation}
    \frac{\mathrm{d} Q_t}{\mathrm{d} t} = -\left(\mu_k + \frac12\sigma_k^2\right) - \sigma_k\eta(t).
    \label{eq:langevin}
\end{equation}
To derive \eqref{eq:langevin}, we must use It\^o's lemma which can be thought of as the analogue of the chain rule for the calculus of stochastic variables~\cite{andersson21}.

The next step is to derive the so-called Fokker-Planck equation (FPE) associated with the stochastic Langevin equation above.
The FPE is a deterministic partial differential equation describing the time-evolution of the probability density function $f_Q$ of the random variable $Q_t$.
It can be shown (see, e.g.,~\cite{zwanzig01}) that the FPE associated with \eqref{eq:langevin} is
\begin{equation}
    \frac{\partial f_Q}{\partial t} = \frac{\sigma_k^2}{2} \frac{\partial^2f_Q}{\partial q^2} + \left(\mu_k + \frac12\sigma_k^2\right)\frac{\partial f_Q}{\partial q}.
    \label{eq:FPE}
\end{equation}

The solution to \eqref{eq:FPE} can be obtained by using the ansatz $f_Q(q,t) = \phi(q,t)\exp(st + rq)$, for some constants $s$ and $r$, and a function $\phi(q,t)$ that must be determined.
Substituting this ansatz in the FPE \eqref{eq:FPE} suggests the following choice of constants
\begin{equation}
r = -\frac{\mu_k + \frac12\sigma_k^2}{\sigma_k^2}\quad\text{and}\quad s=-\frac{\left(\mu_k + \frac12\sigma_k^2\right)^2}{2\sigma_k^2}.
\label{eq:FPE_sol1}
\end{equation}
It then follows that $\phi(q,t)$ satisfies the free diffusion equation
\begin{equation}
    \frac{\partial \phi}{\partial t} = \frac{\sigma_k^2}{2} \frac{\partial^2 \phi}{\partial q^2} \quad \implies \quad \phi(q,t) = \frac{1}{\sqrt{2\pi\sigma_k^2t}}e^{-\frac{q^2}{2\sigma_k^2t}},
    \label{eq:FPE_sol2}
\end{equation}
assuming $\phi(q,0)=\delta(q)$, i.e.\ all the floes have the same size.
Combining \eqref{eq:FPE_sol1} and \eqref{eq:FPE_sol2}, we obtain
\begin{equation}
    f_Q(q,t) = \frac{1}{\sqrt{2\pi\sigma_k^2 t}}e^{-\frac{\left(q + \mu_k t + \frac12\sigma_k^2 t\right)^2}{2\sigma_k^2t}},
\end{equation}
which is a normal distribution at each time $t$ with mean $-\mu_k t - \frac12\sigma_k^2 t$ and variance $\sigma_k^2 t$.
The fact that $Q_t=\ln L_t$ is normally distributed implies that the FSD $f_L$ is lognormal with expression
\begin{equation}
    f_L(l,t) = \frac{1}{l\sqrt{2\pi\sigma_k^2 t}}e^{-\frac{\left(\ln l + \mu_k t + \frac12\sigma_k^2 t\right)^2}{2\sigma_k^2t}}.
\end{equation}

This latter result, although theoretically appealing, raises many questions. 
Is the exponential growth/decay model \eqref{eq:Stoc_exp_model} a reasonable starting point for modelling the stochastic dynamics of floe size?
Is the formulation for the growth/decay rate $k(t)$ appropriate?
And if so, how are the parameters of this formulation, i.e.\ $\mu_k$ and $\sigma_k$, related to the physical properties of the underlying system?
All these questions certainly need to be addressed, but this is beyond the scope of the current paper.
Our goal was to show that the lognormal parametric formulation for the FSD should be recognised as a viable candidate.
We demonstrated that the lognormal FSD naturally emerges from (i) direct wave-induced breakup simulations in a hydroelastic deterministic model~\cite{Mokus2021}, (ii) analysis of field data~\cite{Lu2008} and (iii) a simple stochastic dynamical system~\cite{andersson21}.

We note that using a stochastic dynamical system to model the time evolution of the FSD has been considered before by Herman~\cite{herman10}, who proposed a generalised Lotka-Volterra system of stochastic equations to model conceptually the coupled dynamics of a finite number of floe sizes. 
The resulting steady-state distribution is an inverse-gamma distribution, which the author showed fits reasonably well to field FSD data.
The stochastic framework has also been considered by Toppaladoddi \& Wettlaufer~\cite{toppaladoddi_wettlaufer15} to model the dynamics of the ice thickness distribution (ITD). 
They proposed a way to integrate a FPE associated with a stochastic ice growth process into an ITD transport equation analogous to \eqref{eq:transp1}.
This approach offers a possible way forward for integrating stochasticity in the evolution equation of the FSD, although the key question remains: how can the parameters be derived from the underlying physical mechanisms governing FSD dynamics?

\section{Conclusion}
\label{sec6}

In the last decade, the sea ice modelling community has (i) recognised the importance of incorporating the FSD into large scale sea ice models and (ii) made significant progress to advance the theoretical understanding of the different physical processes affecting the evolution of the FSD and unify them into a self-consistent modelling framework.
The process that has likely received the most attention is wave-induced sea ice breakup, mainly motivated by understanding the impact of the FSD on wave transport in the MIZ.
At the core of most of these studies is an assumption that the FSD follows a power law (or Pareto distribution), a conjecture rooted in early field observations of the FSD. 
Recent evidence from both modelling and observational studies cast doubts on this conjecture.

We proposed a new paradigm for describing the FSD parametrically in terms of the probability density function of a lognormal distribution.
We offered multiple lines of evidence demonstrating the lognormal FSD is an appropriate formulation.
\begin{enumerate}
    \item Simulations of repeated wave-induced breakup using a deterministic hydro-elastic linear model show the resulting FSD is very well fitted by a lognormal distribution for a broad range of wave and ice conditions.
    \item We revisited the analysis of an observational dataset of floe sizes in the Southern Ocean and found the lognormal fitted the FSD better than all the distributions (including the power law) considered in the original analysis.
    \item A simple stochastic mechanism based on the exponential growth/decay dynamics of floe size with random noise is shown to be associated with a time-evolving lognormal probability density function.
\end{enumerate}
Put together, the evidence suggests at the very least that the lognormal FSD needs to be considered as a potential alternative to the power law in future observational and modelling studies.
The stochastic model presented here was only aimed at making a conceptual/qualitative argument for the emergence of the lognormal FSD.
This approach remains largely unexplored, however, and future studies should attempt to make connections between the physical processes affecting the evolution of the FSD and the parameters of the stochastic models.

\vskip6pt

\enlargethispage{20pt}


\textbf{Data accessibility}. The FSD data used to conduct the analysis in \textsection\ref{sec5} were obtained directly from the authors of~\cite{Lu2008}. The source code used to generate the figure and conduct the statistical tests can be obtained from https://doi.org/10.6084/m9.figshare.19416902.v1.

\textbf{Authors' contributions}. FM planned the scope of the manuscript and led the writing effort. NM conducted the analysis of the FSD field data in \textsection\ref{sec5} and contributing some writing in \textsection\ref{sec:sec4} and \ref{sec5}.


\textbf{Funding}. The work was funded through the Marsden Fund project \emph{Breaking the ice: process-informed modelling of sea ice erosion due to ocean wave interactions} (18-UOO-216) administered by Royal Society Te Ap\={a}rangi. 
FM also acknowledges financial support from the New Zealand's Antarctic Science Platform (Project 4) and the Marsden Fund project 20-UOO-173.
NM is supported by a University of Otago doctoral scholarship.

\textbf{Acknowledgement}. We thank Prof.\ Peng Lu (Dalian University of Technology) for kindly sharing their floe size field data.


\bibliographystyle{plain} 

\end{document}